\documentclass{article}

\usepackage[preprint]{neurips_2026}

\usepackage[utf8]{inputenc} 
\usepackage[T1]{fontenc}    
\usepackage{hyperref}       
\usepackage{url}            
\usepackage{booktabs}       
\usepackage{amsfonts}       
\usepackage{nicefrac}       
\usepackage{microtype}      
\usepackage{xcolor}         

\usepackage{amsmath,amssymb,amsfonts}
\usepackage{graphicx}
\usepackage{subcaption}
\usepackage{xspace}
\usepackage{algorithm}
\usepackage{algorithmic}
\usepackage{multirow}
\usepackage{enumitem}
\usepackage{soul}
\usepackage{wrapfig} 

\definecolor{lowratebg}{HTML}{F2F8FE}
\definecolor{midratebg}{HTML}{F3FAF0}
\definecolor{highratebg}{HTML}{FFF4E6}

\newcommand{\rateswatch}[1]{%
    \protect\raisebox{-0.15ex}{%
      \fcolorbox{gray!45}{#1}{\rule{0pt}{0.5em}\hspace{0.5em}}%
    }%
  }

\newcommand{\method}{\textbf{InnVC}\xspace}

\title{High-Fidelity Video Compression based on Invertible Neural Transform and Implicit Conditioning}

\author{%
  \textbf{Siyue~Teng}\thanks{Equal contribution.}, 
  \textbf{Ho~Man~Kwan}\footnotemark[1], 
  \textbf{Yuxuan~Jiang}, 
  \textbf{Fan~Zhang}, 
  \textbf{David~Bull} \\
  Visual Information Lab, University of Bristol, UK\\
  \texttt{\{siyue.teng, hm.kwan, yuxuan.jiang, fan.zhang, dave.bull\}@bristol.ac.uk} \\
}

\begin{document}

\maketitle

\begin{abstract}
Learning-based video compression has recently achieved competitive rate-distortion performance compared to conventional video codecs. However, most existing methods rely on non-invertible analysis-synthesis transforms, with reconstruction quality subject to both quantization and transform approximation errors. This limitation becomes particularly restrictive at higher quality points, where quantization errors are small and transform-induced distortion dominates. To address this, we propose \method, an \textbf{I}nvertible \textbf{n}eural \textbf{n}etwork based \textbf{V}ideo \textbf{C}odec for wide-range and high-fidelity compression. The core idea is to preserve an invertible main transform path prior to quantization, while injecting content-adaptive context through a compact implicit conditioning field. This decouples strongly correlated video content from harder-to-model fine details, allowing different components to specialize in complementary reconstruction tasks for more efficient compression. To further improve compressibility, we introduce a scheduled masking strategy that progressively concentrates informative content into fewer latent channels for more effective entropy coding. 
Experiments on the UVG and MCL-JCV benchmarks show that \method achieves strong compression performance over a broad quality range, being particularly effective in the high-quality regime, yielding BD-rate reductions of 21.66\% in PSNR and 46.06\% in MS-SSIM relative to x265 on UVG. To the best of our knowledge, \method is the first neural video codec covers operating poins from low bitrate to high fidelity within a single architecture scale, spanning more than \textbf{20 dB} in PSNR.
\end{abstract}

\section{Introduction}

Video compression underpins modern visual communication and media delivery, enabling applications such as streaming, video conferencing, cloud gaming, and immersive media~\cite{bull2021intelligent}. The continual demand for increased spatial resolution, frame rate, dynamic range, and visual fidelity further increases the demand for efficient, high-quality video coding~\cite{cisco2022vni,teng2024benchmarking}. For decades, this problem has been addressed by traditional video compression standards, including H.264/AVC~\cite{wiegand2003overview}, H.265/HEVC~\cite{sullivan2012overview}, H.266/VVC~\cite{bross2021overview}, and AV1~\cite{han2020technical}. These methods combine prediction, transform coding, quantization and entropy coding, refined with sophisticated tools~\cite{chen2020overview,hamidouche2022versatile} that achieve strong performance~\cite{zhao2022aom,abdoli2024video}. By controlling quantization strength and coding modes, such approaches are effective across a wide quality range, from high compression to high-fidelity configurations.

Neural video compression has recently emerged as a strong alternative to conventional hybrid video codecs and has demonstrated competitive, and in some cases, superior rate-distortion performance~\cite{Gao2026advances}. Early learned video compression methods were based on autoencoder-based architectures, where learnable analysis and synthesis transforms are combined with temporal prediction, context modeling, and entropy models~\cite{lu2019dvc,li2021deep} to enable efficient video coding. In parallel, recent research has focused on video representations using an overfitted model, such as an implicit neural representation (INR)~\cite{chen2021nerv,kwan2023hinerv,gao2025givic}. Unlike autoencoder-based codecs, that encode videos into latent representations, INR-based methods model a video as a continuous neural function parameterized by network weights, which can offer compact sequence-specific representations and flexible scalability.

A key challenge in learned video compression is to maintain strong performance across a wide bitrate/quality range. Low-bitrate coding requires the effective exploitation of spatio-temporal redundancy, whereas high-bitrate coding demands a transform path with minimal intrinsic information loss. Although neural video compression has substantially improved coding efficiency, existing approaches have primarily focused on the low- to medium-bitrate regime. While some works attempt to broaden the supported quality range through carefully designed training and rate-control strategies~\cite{li2024neural,zhang2024learned}, these approaches have yet to reach the high-fidelity or near-lossless range consistently.
As a result, wide quality coverage, especially at the high-quality end, remains underexplored in current learned video codecs.

Existing learned video compression methods generally attain higher reconstruction quality by using low quantization steps. However, reconstruction distortion stems not only from quantization error but also from inherent defects of learned transforms. Typical learned transforms, such as autoencoders, fail to perfectly recover input images and video frames, thereby introducing intrinsic distortion. This bottleneck becomes particularly prominent in high-quality coding scenarios, where, with fine-grained quantization, transform-induced distortion becomes predominant, severely limiting overall reconstruction quality. This motivates us to develop a neural codec that maintains invertibility before quantization, which is consistent with conventional codecs, through optimized transform design rather than introducing additional architectural complexity.

\begin{figure}[t]
  \centering
  \includegraphics[width=1.0\linewidth]{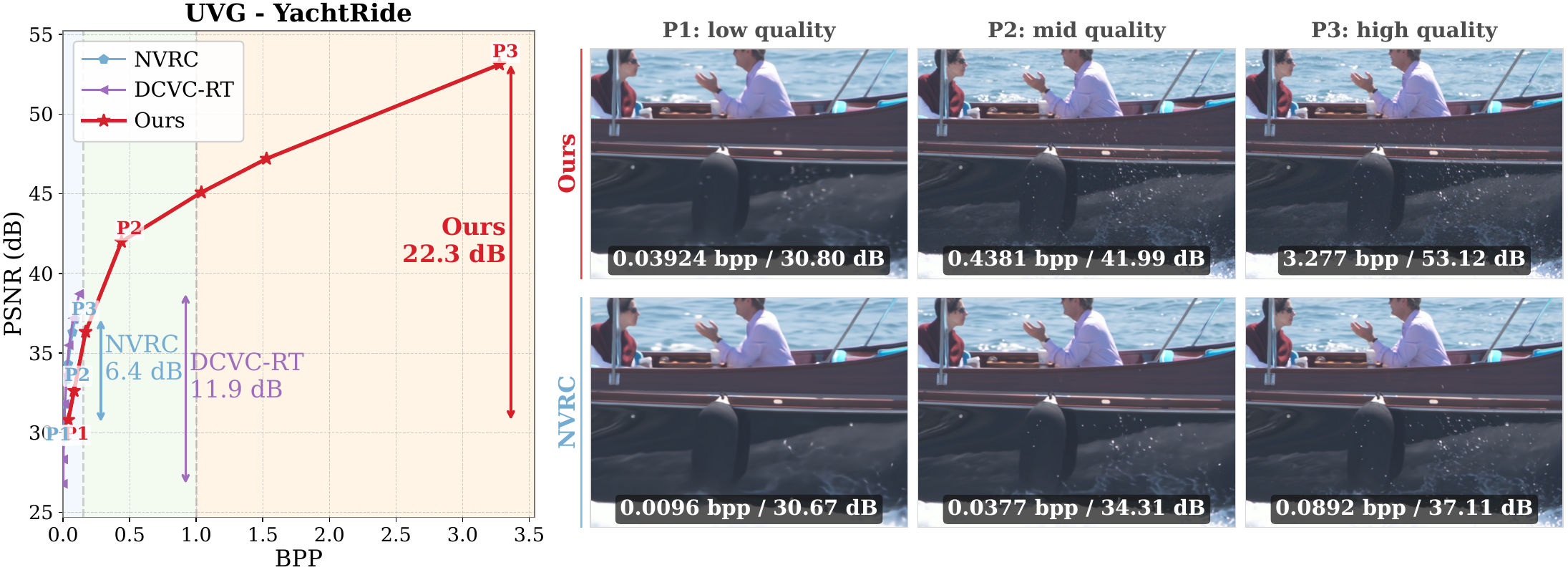}
  \caption{\textbf{Left:} RD curve of the YachtRide sequence in UVG. The shaded RD background provides a qualitative visual guide to
  \rateswatch{lowratebg} lower,
  \rateswatch{midratebg} medium, and
  \rateswatch{highratebg} higher BPP regions from left to right.
  \textbf{Right:} Visual comparisons sampled from each method's attainable quality range, from low to middle to highest available quality. Ours spans a significantly wider quality range.}
  \label{Fig:teaser}
\end{figure}

Invertible neural networks (INN) provide a natural foundation for this goal, since they are based on bijective transformations with exact inverses by design. Some related work has already shown promise in image compression \cite{xie2021enhanced,cai2024i2c,tu2025multi,gao2025approximately}; however, their application to video compression has been limited \cite{montajabi2023invertible,guo2025exploring}. Directly applying a bijective transform to video does not automatically yield a compression-friendly representation, because a bijective transform preserves dimensionality - spatial squeezing mainly reorganizes information into the channel dimension - rather than removing redundancy. Hence, for videos, where strong spatio-temporal correlation is prevalent,  additional structure may still be needed to make the latent representation efficient for compression. 

In this context, we propose a multi-stage invertible backbone with a compact implicit conditioning field that provides content-adaptive modulation. The key idea is to decouple strongly correlated spatio-temporal content from less correlated fine details. The spatio-temporal redundancy is captured via the implicit conditioning field and injected into the invertible transform with multi-scale modulation. Meanwhile, the hard-to-capture details are modeled directly by the invertible backbone. In this way, the model preserves a truly invertible transform path prior to quantization, while still exploiting rich video-dependent context for efficient compression. To further make the resulting high-dimensional latent representation compression-friendly, we avoid squeezing all information into a narrow feature subset. Instead, we induce a regularization technique that leverages masking to enhance the high-dimensional latent representation, making it more amenable to entropy modeling and more suitable for channel-dependent quantization. Figure~\ref{Fig:teaser} highlights the two main properties of the proposed design. First, removing the INN backbone causes the reconstruction quality to saturate even when the bitrate continues to increase, showing that the invertible transform is essential for reaching the high-fidelity regime. Second, compared with existing codecs, \method covers a much broader rate-distortion range. The main contributions of this work are summarized as follows.

\begin{itemize}[leftmargin=*]
  \item We propose \method, an invertible neural video compression method for \textbf{wide-range and high-fidelity compression}. It employs a \textbf{multi-stage invertible backbone to remove transform approximation error} from the main coding path, thereby alleviating a key bottleneck in the high-quality regime.

  \item We introduce a \textbf{multi-scale modulation design} that couples the invertible backbone with an \textbf{implicit conditioning field}. This design decouples strongly correlated spatio-temporal content from less correlated fine details, improving redundancy modeling efficiently.

  \item We propose a \textbf{scheduled masking strategy} combined with a \textbf{channel autoregressive model} that improves latent channel organization for more effective entropy coding.
\end{itemize}

Experiments on the UVG and MCL-JCV benchmarks show that \method achieves competitive coding performance across a broad quality range and performs particularly well in the high-quality regime, consistently outperforming x265 on both datasets under PSNR and MS-SSIM, achieving average BD-rate reductions of 24.90\% on UVG and 22.38\% on MCL-JCV in terms of PSNR.

\section{Related Work}

\noindent\textbf{Learned Video Compression.} Autoencoder-based transform architectures have been widely adopted in learned video codecs, including early residual-coding frameworks~\cite{lu2019dvc,agustsson2020scale,hu2021fvc,hu2020improving} and more recent conditional-coding variants~\cite{li2021deep,li2022hybrid,li2023neural,li2024neural,jia2025towards}. These methods improve coding efficiency through learned motion compensation, temporal priors, context propagation, and stronger entropy models~\cite{sheng2022temporal,xiang2023mimt,tang2025neural,jiang2025ecvc,li2022hybrid}. Some recent works further broaden the quality range of a single model through quantization scaling, parameter sampling, or rate-control mechanisms~\cite{li2024neural,jia2025towards}. Despite their diversity, these approaches still rely on lossy analysis-synthesis transforms, so reconstruction quality is affected by both quantization and transform approximation errors.

A complementary line of work represents each video with an overfitted neural representation, typically an INR~\cite{chen2021nerv,chen2023hnerv,bai2023ps,lee2023ffnerv}. These methods provide compact, sequence-specific representations with strong reconstruction quality and efficient decoding, and recent works further improve parameter compression through better quantization and entropy coding~\cite{shi2025quantizing}. In particular, multi-scale spatio-temporal grids and related structured representations compactly represent a large amount of correlated video content~\cite{lee2023ffnerv,kwan2023hinerv,maiya2023nirvana,kwan2024immersive,kwan2024nvrc}. However, bitrate and quality control in this family are often realized through model scaling or parameter compression, so different operating points typically correspond to representations of different sizes, and the supported quality range of a single configuration is often limited.

\noindent\textbf{Invertible Network Based Compression.} Invertible neural networks define bijective mappings between the input and latent spaces, making the transform stage information-preserving prior to quantization. They are most widely studied in normalizing flows~\cite{kobyzev2020normalizing}, but have also been explored as learnable transforms for compression: coupling-based invertible architectures were popularized by NICE~\cite{dinh2014nice} and RealNVP~\cite{dinh2016density}, and several later works extended these to image compression. In particular,~\cite{helminger2020lossy} first applied normalizing flows to lossy image compression, while later methods improved practical rate-distortion performance through augmented flows, channel squeeze, feature enhancement, variable-rate invertible transforms, or approximately invertible formulations~\cite{ho2021anfic,xie2021enhanced,cai2024i2c,tu2025multi,gao2025approximately}. Related ideas have also been explored with trainable wavelet-like invertible transforms for lossy and lossless image compression~\cite{ma2020end,ma2019iwave}. Collectively, these works show that invertibility can reduce transform-side information loss, but also that exact reversibility alone does not necessarily produce a latent representation that is easy to compress.

Only a limited number of works have extended invertible compression to learned video coding. CANF-VC~\cite{ho2022canf} uses conditional augmented normalizing flows for conditional video coding, where invertibility mainly serves probabilistic modeling. Other recent works explore invertible encoding more directly for video compression~\cite{montajabi2023invertible,guo2025exploring}, but still rely on auxiliary designs such as channel squeeze or conditional motion coding to make the representation practically compressible. In contrast, our work places exact invertibility of the main transform path prior to quantization at the center of the codec design, and combines it with an implicit conditioning field to make such a transform practical for high-fidelity video compression.

\section{Method}

\subsection{Preliminaries}
\begin{figure}[t]
  \centering
  \includegraphics[width=1.0\linewidth]{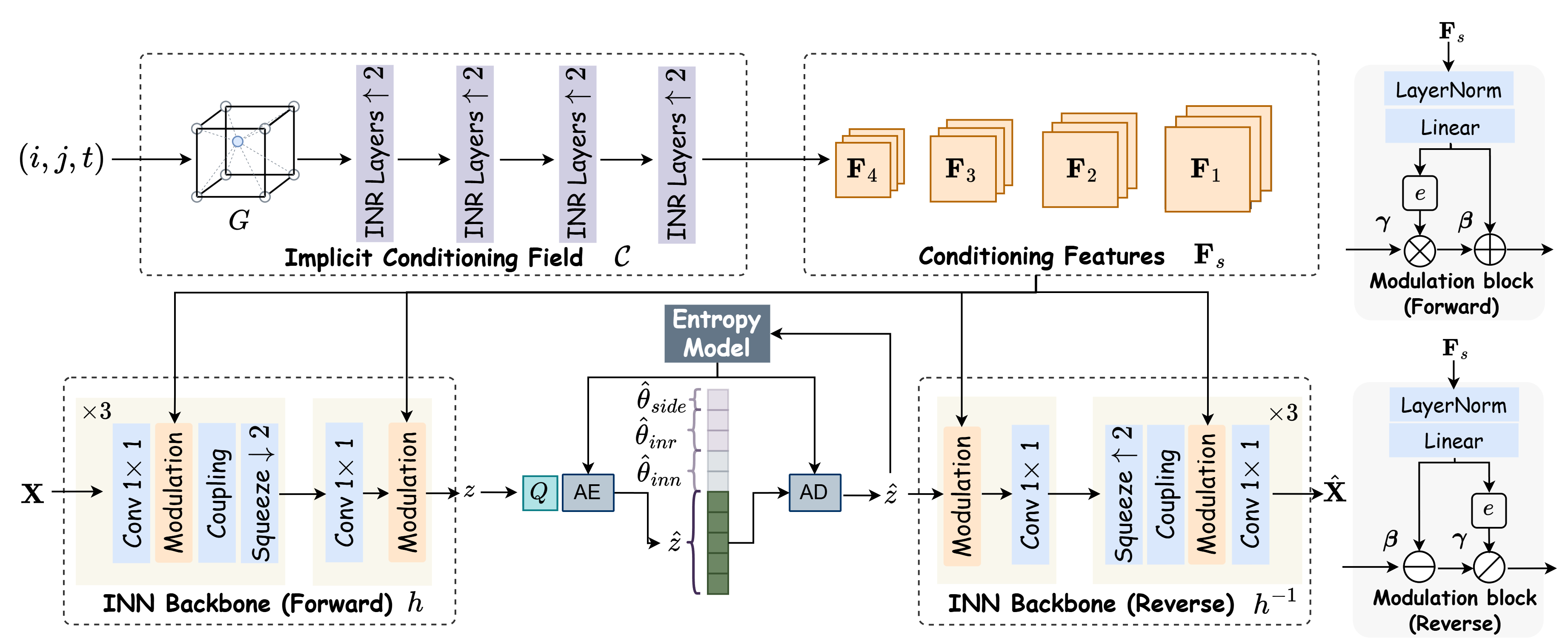}
  \caption{\textbf{Left:} Overview of the proposed InnVC framework. The INN backbone encodes and decodes the input video patches, and the implicit conditioning field provides multi-scale features for modulation. $Q$ denotes quantization, and AE/AD denote arithmetic encoding and arithmetic decoding, respectively. \textbf{Right:} The modulation mechanism, where the conditioning feature is transformed into channel-wise scale and shift parameters and applied to the backbone feature.}
  \label{Fig:framework}
\end{figure}

\noindent\textbf{Implicit neural representations.}
Implicit neural representations (INRs) model a visual signal as a coordinate-conditioned neural function parameterized by a compact set of learnable parameters~\cite{chen2021nerv}. In video representation, an INR takes spatio-temporal coordinates as input and predicts the corresponding content, often together with structured feature grids to improve representation capacity and decoding efficiency~\cite{lee2023ffnerv,kwan2023hinerv}. In this work, we adopt this paradigm only for conditioning: the INR-style module generates multi-scale modulation features for the invertible backbone rather than reconstructing video pixels directly.

\noindent\textbf{Invertible neural networks.}
An invertible neural network (INN) defines a bijective mapping $h:\mathbf{x}\leftrightarrow\mathbf{y}$~\cite{dinh2014nice,dinh2016density}, such that both the forward transform and its inverse are tractable:
\begin{equation}
\mathbf{y}=h(\mathbf{x}), \qquad \mathbf{x}=h^{-1}(\mathbf{y}).
\end{equation}
A common invertible design is based on affine coupling layers~\cite{dinh2016density}, where the input is partitioned into two parts - with one being transformed conditional on the other. This yields an analytically invertible mapping and forms the basis of many INN architectures used in generative modeling and compression. In this work, we build on this principle to construct an invertible transform backbone for video compression.

\subsection{Framework}
\label{sec:framework}

We propose \method, an instance-adaptive video codec based on invertible neural networks, as illustrated in \autoref{Fig:framework}. The framework consists of two core components: an \textbf{INN-based transform backbone} (\autoref{sec:invertible_backbone}) and an \textbf{implicit conditioning field} (\autoref{sec:conditioning_field}). The conditioning field produces multi-scale modulation features, while the INN backbone performs the main analysis and synthesis transforms.

Given an input video $\mathbf{V}\in\mathbb{R}^{T\times H\times W\times 3}$, we randomly sample overlapping patches and denote the sampled patch tensor as $\mathbf{X}$. The INN backbone maps each patch to a latent representation through a multi-stage forward transform conditioned on the modulation features $\{\mathbf{F}_s\}_{s=1}^{S}$, while the reconstruction $\hat{\mathbf{X}}$ is obtained from the quantized latent $\hat{\mathbf{z}}$ through the inverse transform:
\begin{equation}
\mathbf{z}=h(\mathbf{X};\{\mathbf{F}_s\}_{s=1}^{S}),
\qquad
\hat{\mathbf{X}}=h^{-1}(\hat{\mathbf{z}};\{\mathbf{F}_s\}_{s=1}^{S}).
\end{equation}

The modulation features are generated by the implicit conditioning field from the patch coordinate and a compact learned feature grid $\mathbf{G}$. Specifically, for a patch located at the spatio-temporal coordinate $(i,j,t)$, the conditioning field takes $(i,j,t)$ together with the quantized feature grid parameters $\hat{\mathbf{G}}$ as input and outputs the multi-scale features:
\begin{equation}
\{\mathbf{F}_s\}_{s=1}^{S}=\mathcal{C}\!\left((i,j,t),\hat{\mathbf{G}}\right).
\end{equation}
Here, we adopt grid-based INRs - influenced by recent works that have shown that structured feature grids provide stronger representation capacity and decoding efficiency than purely coordinate-driven designs~\cite{ladune2023cool,lee2023ffnerv,kwan2023hinerv}.

The overall decoding process is therefore determined by the quantized conditioning representation $\hat{\mathbf{G}}$ and the quantized INN latent $\hat{\mathbf{z}}$. Unlike existing INR-based codecs that rely on separate encoder-decoder transforms~\cite{chen2023hnerv}, our method uses an invertible backbone as the main transform path, with an INR-style module only for conditioning. This design combines exact invertibility before quantization with strong content-adaptive modulation. Since a na\"ive invertible transform produces a channel-heavy latent that is difficult to compress efficiently, we further introduce a scheduled masking strategy to organize the latent channels into a more compression-friendly structure; details are provided in \autoref{sec:scheduling_mask}.

\subsection{Implicit Conditioning Field}
\label{sec:conditioning_field}

We model the implicit conditioning field $\mathcal{C}(\cdot)$ using a compact feature representation together with a multi-scale decoder $\{\mathcal{D}_s\}_{s=1}^{S}$. The base representation is parameterized by a learnable feature grid $\mathbf{G}$~\cite{lee2023ffnerv,ladune2023cool,kwan2023hinerv} of size $T_G \times H_G \times W_G \times C_G$, whose spatio-temporal resolution is only a fraction of that of the full video. Given a video patch located at spatio-temporal coordinate $(i,j,t)$, we extract a patch-specific base feature from $\hat{\mathbf{G}}$ using trilinear interpolation, and then feed it to the decoder to progressively generate multi-scale modulation features:
\begin{equation}
\mathbf{F}_{\mathrm{base}} = \operatorname{Interp}((i,j,t), \hat{\mathbf{G}}), \qquad
\mathbf{F}_1 = \mathcal{D}_1(\mathbf{F}_{\mathrm{base}}), \qquad
\mathbf{F}_s = \mathcal{D}_s(\mathbf{F}_{s-1}),\; s=2,\dots,S.
\end{equation}

The resulting features are configured to match the resolutions of the corresponding stages in the invertible backbone: the decoder progressively upsamples its features, while the backbone reduces spatial resolution through successive squeeze operations. Following prior INR-based video representations~\cite{lee2023ffnerv,ladune2023cool,kwan2023hinerv}, we implement $\mathbf{G}$ as a multi-resolution feature grid to improve representation capacity while remaining compact. The overall design of the conditioning field is mainly based on~\cite{kwan2023hinerv}. Additional implementation details are provided in the supplementary material.

\subsection{Modulated Invertible Backbone}
\label{sec:invertible_backbone}

The forward and inverse paths of the invertible backbone serve as the analysis and synthesis transforms, conditioned on the multi-scale features $\{\mathbf{F}_s\}_{s=1}^{S}$ produced by the implicit conditioning field. The backbone maps an input patch to a latent representation through $S$ stages of reversible transforms operating at progressively lower spatial resolutions. In each stage, let $\mathbf{z}^{(0)}=\mathbf{X}$ and $\mathbf{z}^{(S)}=\mathbf{z}$. The backbone contains $S-1$ multi-resolution stages followed by a final refinement stage. Each multi-resolution stage applies $\mathrm{Conv}_{1\times1}$, feature modulation, affine coupling, and spatial squeeze in sequence, while the final stage omits the coupling and squeeze operation.

As successive squeeze operations reduce spatial resolution and increase channel capacity, the representation becomes increasingly channel-heavy at deeper stages. Since the transform is invertible, information is preserved prior to quantization, but the resulting latent is also more difficult to compress efficiently. The conditioning field alleviates this by injecting structured spatio-temporal context at each stage.

\textbf{Invertible Convolution.}
For channel mixing, we adopt a learned orthogonal $\mathrm{Conv}_{1\times1}$ layer~\cite{kingma2018glow}. The orthogonal weight matrix is parameterized through the Cayley transform~\cite{cayley1846quelques} of a skew-symmetric matrix, which guarantees invertibility by construction and avoids explicit matrix inversion during decoding.

\textbf{Modulation Layer.}
The conditioning feature $\mathbf{F}_s$ modulates the output of $\mathrm{Conv}_{1\times1}$. Specifically, $\mathbf{F}_s$ is processed by LayerNorm and a linear layer to produce channel-wise scale and shift parameters $(\boldsymbol{\gamma}_s,\boldsymbol{\beta}_s)$, which are then applied through an affine modulation operation.

\textbf{Coupling Layer.}
Following standard invertible architectures~\cite{dinh2016density,kingma2018glow}, we adopt an affine coupling layer as the nonlinear transform at each stage. Concretely, we use a two-sided coupling design in which both channel partitions are updated sequentially using two lightweight subnetworks. In our implementation, the subnetworks are instantiated with ConvNeXt blocks~\cite{liu2022convnet} for parameter efficiency.

For stages $s=1,\dots,S-1$, the forward transform applies an invertible $\mathrm{Conv}_{1\times1}$ layer, feature modulation, affine coupling, and squeeze, where
$(\boldsymbol{\beta}_s,\boldsymbol{\gamma}_s)=\operatorname{Linear}(\operatorname{LayerNorm}(\mathbf{F}_s))$:
\begin{equation}
\bar{\mathbf{z}}^{(s)} = \mathrm{Conv}_{1\times1}\!\left(\mathbf{z}^{(s-1)}\right),
\qquad
\mathbf{q}^{(s)} = \bar{\mathbf{z}}^{(s)} \odot \exp(\boldsymbol{\gamma}_s) + \boldsymbol{\beta}_s,
\end{equation}
\begin{equation}
\mathbf{z}^{(s)} = \operatorname{Squeeze}\!\left(\operatorname{Coupling}\!\left(\mathbf{q}^{(s)}\right)\right).
\end{equation}
The final stage serves as a refinement stage at the deepest resolution and omits the coupling and squeeze operations.
\begin{equation}
\bar{\mathbf{z}}^{(S)} = \mathrm{Conv}_{1\times1}\!\left(\mathbf{z}^{(S-1)}\right),
\qquad
\mathbf{z}^{(S)} = \bar{\mathbf{z}}^{(S)} \odot \exp(\boldsymbol{\gamma}_S) + \boldsymbol{\beta}_S.
\end{equation}

Given the decoded conditioning features ${\mathbf{F}s}{s=1}^{S}$, the squeeze operation, the orthogonal $\mathrm{Conv}_{1\times1}$ layer, the affine modulation, and the coupling transform are all bijective. Therefore, each stage remains invertible, and so does the overall mapping $h(\cdot;\{\mathbf{F}_s\}_{s=1}^{S})$.

\subsection{Scheduled Masking for Channel Organization}
\label{sec:scheduling_mask}

Although the invertible backbone preserves information, the resulting latent representation $\mathbf{z}$ is high-dimensional and heavily distributed across channels. To make this representation more amenable to compression, we introduce a scheduled masking strategy that progressively regularizes the latent channels, encouraging more informative content to concentrate in the earlier channels. Additional implementation details are provided in the supplementary material.

Let $\mathbf{z}\in\mathbb{R}^{T\times H'\times W'\times C}$ denote the latent produced by the INN encoder. During training, we partition the channel dimension into ordered groups and assign each group $g$ a time-dependent activation schedule $\rho_g(\tau)\in[0,1]$, where $\tau\in[0,1]$ denotes the normalized training progress and $\rho_g(\tau)$ controls the probability that group $g$ becomes active at training progress $\tau$. At the beginning of training, all latent channels are masked, so reconstruction mainly relies on the conditioning branch. This prevents the invertible branch from dominating reconstruction too early and allows the conditioning branch to first learn a strong content representation. As training proceeds, channel groups are progressively activated from early to late. Moreover, once a later group is activated, all preceding groups are also activated. This cumulative activation causes earlier channels to participate in training more frequently and therefore encourages informative content to concentrate in the leading channels, in contrast to standard dropout, which mainly serves as a regularizer rather than explicitly learning ordered representations~\cite{rippel2014learning,srivastava2014dropout}. As a result, the latent channels become progressively organized by importance, making the representation easier to model with an entropy coder and better suited to channel-dependent quantization.

\subsection{Entropy Coding and Rate-Distortion Optimization}
\label{sec:training_objective}

\noindent\textbf{Entropy coding.}
All transmittable variables in \method are quantized and entropy coded, including the primary latent $\hat{\mathbf{z}}$ produced by the invertible backbone, the parameters of the implicit conditioning field and the invertible backbone, denoted by $\hat{\theta}_{\mathrm{inr}}$ and $\hat{\theta}_{\mathrm{inn}}$, and the side parameters of the quantization and entropy models, denoted by $\hat{\theta}_{\mathrm{side}}$. The quantized feature grid $\hat{\mathbf{G}}$ is included in $\hat{\theta}_{\mathrm{inr}}$.

For the INN latent $\hat{\mathbf{z}}$, we adopt a channel autoregressive entropy model~\cite{minnen2020channel} with conditional Gaussian distributions. The latent channels are organized into ordered groups, and earlier groups are encoded and decoded before later ones, so that they can serve as context for subsequent groups. We implement this channel autoregressive model using a lightweight masked convolutional network~\cite{aaron2016pixel}. Compared with commonly used spatial autoregressive models, the channel-wise design is better aligned with the channel-organized latent produced by the invertible backbone and enables more efficient entropy coding.

For the remaining parameters, including $\hat{\theta}_{\mathrm{inr}}$, $\hat{\theta}_{\mathrm{inn}}$, and $\hat{\theta}_{\mathrm{side}}$, we follow the compression framework of NVRC~\cite{kwan2024nvrc}. Additional details of the entropy models are provided in the supplementary material.

At encoding time, the input patch is first mapped to the latent representation $\mathbf{z}$ using the invertible backbone conditioned on the decoded features from the implicit conditioning field. The latent is then quantized to obtain $\hat{\mathbf{z}}$, which is entropy coded together with the quantized model parameters into the final bitstream.

\textbf{Rate-distortion optimization.}
\method is optimized in an instance-adaptive manner for each input video under a standard Lagrangian rate-distortion objective:
\begin{equation}
\mathcal{L} = D(\mathbf{X},\hat{\mathbf{X}}) + \lambda R,
\end{equation}
where $D(\mathbf{X},\hat{\mathbf{X}})$ measures the reconstruction distortion, $R$ denotes the estimated total bitrate~\cite{balle2017end}, and $\lambda$ controls the rate-distortion trade-off. The total bitrate is given by $R = R_{\mathbf{z}} + R_{\mathrm{inn}} + R_{\mathrm{inr}} + R_{\mathrm{side}}$, where $R_{\mathbf{z}}$, $R_{\mathrm{inn}}$, $R_{\mathrm{inr}}$, and $R_{\mathrm{side}}$ denote the rates of $\hat{\mathbf{z}}$, $\hat{\theta}_{\mathrm{inn}}$, $\hat{\theta}_{\mathrm{inr}}$, and $\hat{\theta}_{\mathrm{side}}$, respectively.

During training, quantization is approximated by a differentiable soft-rounding scheme together with a straight-through estimator~\cite{kim2024c3,agustsson2020universally,bengio2013estimating}, allowing gradients to propagate through both the latent representation and the conditioning parameters. At inference time, actual quantization is used for entropy coding and exact decoding.


\section{Experimental Setup}

\noindent\textbf{Datasets.}
We evaluate \method on two widely used video compression benchmarks, UVG~\cite{mercat2020uvg} and MCL-JCV~\cite{wang2016mcl}. UVG contains 7 videos at $1920\times1080$ resolution, each with either 300 or 600 frames. MCL-JCV contains 30 video clips, each with 120 to 150 frames at the same resolution. 

\noindent\textbf{Baselines.}
We compare \method against seven representative baselines from three categories. Conventional codecs: x265~\cite{x265} with the \textit{veryslow} preset, and HM 18.0~\cite{hm18}, both under the Random Access (RA) configuration. We also include Apple ProRes~\cite{apple2022prores} as a practical high-quality intra/intermediate codec reference, since its frame-independent coding structure provides a useful comparison point in the high-fidelity regime. Learned video codecs: DCVC-FM~\cite{li2024neural} and DCVC-RT~\cite{jia2025towards}. INR-based codecs: HiNeRV~\cite{kwan2023hinerv} and NVRC~\cite{kwan2024nvrc}. This setting enables comparison with both classical codecs and recent neural video compression methods.

\noindent\textbf{Evaluation metrics.}
We use Bjøntegaard Delta rate (BD-rate)~\cite{bjontegaard2001calculation} as the primary metric for rate-distortion comparison, with x265 as the anchor. Bitrate is measured in bits per pixel (bpp), while reconstruction quality is evaluated using PSNR and MS-SSIM~\cite{wang2003multiscale}. These metrics are typical in the learned and INR-based video compression literature.

\noindent\textbf{\textbf{Implementation details.}}
Following common practice in INR-based compression, \method is trained independently for each video and each rate point, without large-scale offline pretraining. We use a 4-stage multi-scale ConvNeXt-based conditioning decoder together with a 4-stage invertible backbone. The model is trained for 360 epochs on UVG and 720 epochs on MCL-JCV using randomly sampled $120\times120$ patches with a batch size of 144. Optimization is performed using the Adam optimizer ($\beta_1=0.9$, $\beta_2=0.999$, $\epsilon=10^{-6}$), with a base learning rate of $2\times10^{-3}$, and cosine annealing with warmup. We evaluate rate-distortion trade-offs with $\lambda\in\{1,4,8,16,32,64,96,128\}$. The distortion loss is defined as $0.7\mathcal{L}_1 + 0.3(1-\mathrm{MS\text{-}SSIM})$. Training uses mixed precision, while the INN is computed in FP32 for numerical stability. For the proposed scheduled masking, the 256 INN latent channels are divided into 20 groups and progressively activated from 40\% to 80\% of training. Additional architectural and training details are provided in the supplementary material.

\begin{table}[t]
\centering
\caption{BD-rate results on the UVG~\cite{mercat2020uvg} and MCL-JCV~\cite{wang2016mcl} datasets, with x265 as the anchor. Negative values indicate better compression performance. Results are reported separately for the low-bitrate regime (BPP $<$ 0.6) and the high-bitrate regime (BPP $\geq$ 0.6) to better match overlapping rate ranges for BD-rate calculation. $-$ indicates that BD-rate cannot be reliably computed due to limited overlap between the corresponding R-D curves.}
\label{tab:bdrate}
\resizebox{\linewidth}{!}{
\begin{tabular}{l|ll|cccccccc}
\toprule
Category & Dataset & Metric
            & x265      & NVRC             & HiNeRV        & DCVC-FM   & DCVC-RT       & HM (\textit{RA})      & ProRes        & Ours \\
\midrule
\multirow{4}{*}{Low-bitrate}
& \multirow{2}{*}{UVG}
& PSNR      & 0.0 \%      & \textbf{-72.78\%}  & -38.66\%        & -55.20\%    & -59.86\%        & -34.61\%                & $-$           & -28.14\% \\
& & MS-SSIM & 0.0\%       & \textbf{-83.65\%}  & -62.70\%        & -62.04\%    & -65.60\%        & -35.23\%                & $-$           & -61.50\% \\
\cmidrule(lr){2-11}
& \multirow{2}{*}{MCL-JCV}
& PSNR      & 0.0\%       & -51.61\%              & -23.39\%           & \textbf{-51.79\%}    & -48.49\%        & -37.47\%                & $-$           & -26.74\% \\
& & MS-SSIM & 0.0\%       & \textbf{-66.83\%}              & -44.12\%           & -56.42\%    & -57.09\%        & -38.88\%                & $-$           & -35.11\% \\
\midrule
\multirow{4}{*}{High-bitrate}
& \multirow{2}{*}{UVG}
& PSNR      & 0.0\%       & $-$               & $-$           & $-$       & $-$           & -18.16\%                & 137.64\%        & \textbf{-21.66\%} \\
& & MS-SSIM & 0.0\%       & $-$               & $-$           & $-$       & $-$           & -9.91\%                 & 73.25\%         & \textbf{-46.06\%} \\
\cmidrule(lr){2-11}
& \multirow{2}{*}{MCL-JCV}
& PSNR      & 0.0\%       & $-$               & $-$           & $-$       & $-$           & \textbf{-19.41\%}                & 159.06\%        & -18.01\% \\
& & MS-SSIM & 0.0\%       & $-$               & $-$           & $-$       & $-$           & -10.21\%                & 112.34\%        & \textbf{-28.73\%} \\
\bottomrule
\end{tabular}
}
\end{table}

\begin{figure}[t]
\centering
\begin{subfigure}[t]{0.495\textwidth}
  \centering
  \includegraphics[width=\linewidth]{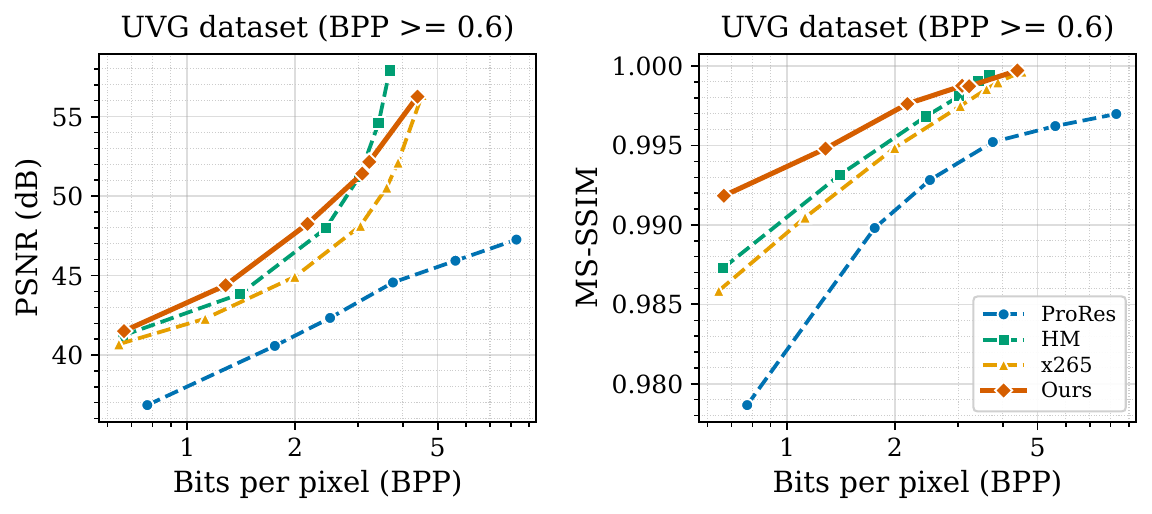}
\end{subfigure}
\hfill
\begin{subfigure}[t]{0.495\textwidth}
  \centering
  \includegraphics[width=\linewidth]{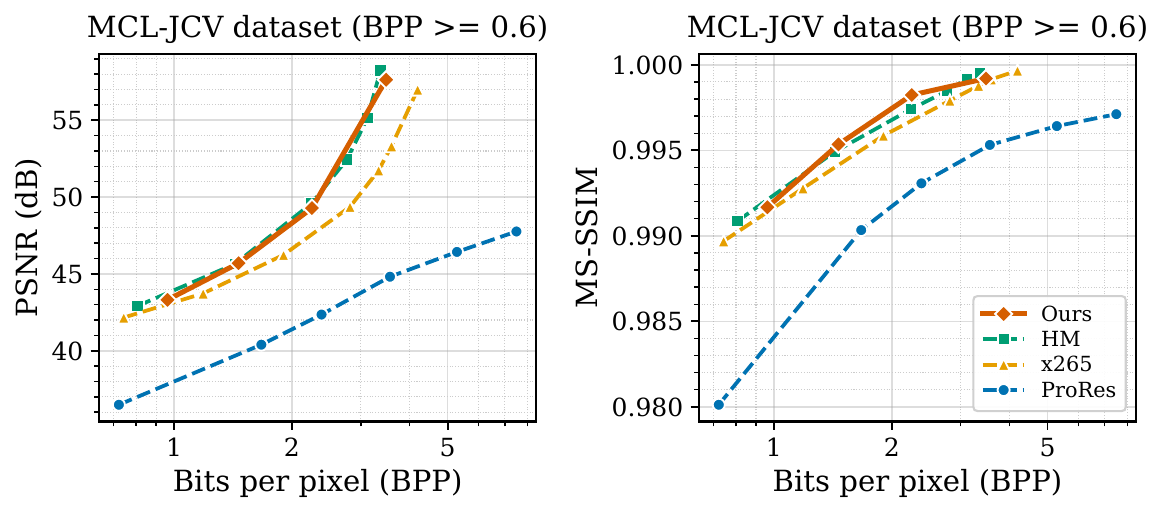}
\end{subfigure}

\vspace{0.5em}

\caption{ Rate-distortion curves on the UVG and MCL-JCV datasets.}
\label{fig:rd_curve}
\end{figure}

\section{Results and Discussion}

\noindent\textbf{\textbf{Rate-distortion performance.}}
\autoref{fig:rd_curve} reports the rate-distortion curves on the UVG and MCL-JCV datasets, and the corresponding BD-rate results are summarized in \autoref{tab:bdrate}. We report results separately in the low-bitrate and high-bitrate regions. The symbol ``$-$'' indicates that the corresponding result is unavailable, either because the method does not target that bitrate range or because the result is not reported by the original paper.

In the high-bitrate region, \method is the only neural codec operating there, yielding 18.01\% and 28.73\% BD-rate savings relative to x265 under PSNR and MS-SSIM on MCL-JCV, respectively. It also outperforms ProRes on both datasets and beats HM (RA) on UVG. In the low-bitrate region, \method remains competitive, with 28.14\% and 26.74\% BD-rate savings over x265 on UVG and MCL-JCV datasets, respectively. Compared with NVRC, \method performs worse at the low-bitrate end, which is consistent with our design choice of using a single model scale across multiple $\lambda$ values while prioritizing broad-range and high-fidelity compression.

\begin{figure}[t]
  \centering
  \includegraphics[width=\linewidth]{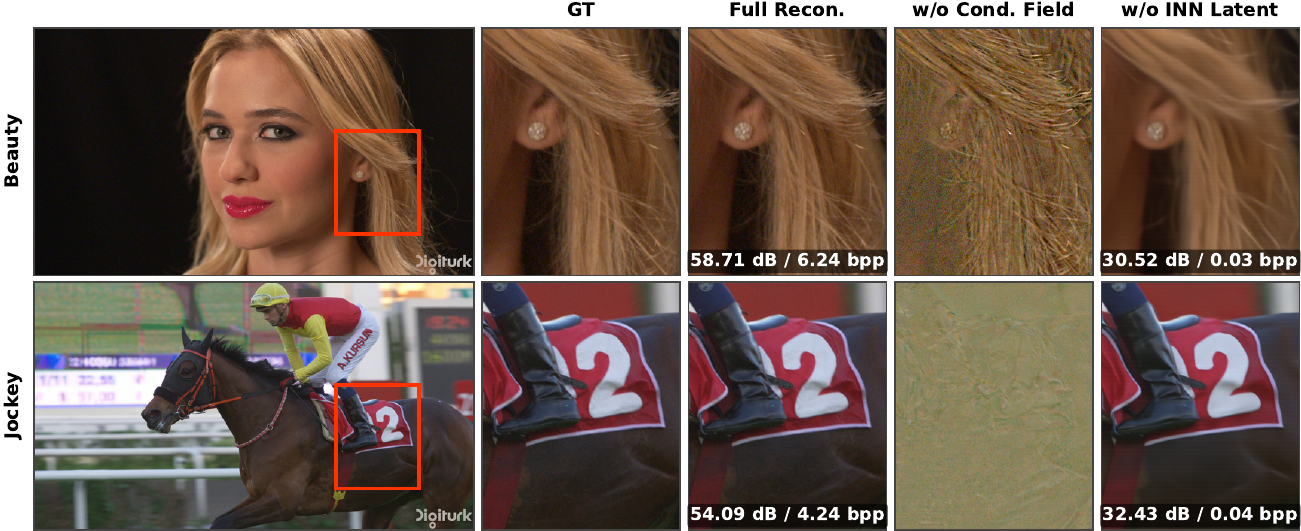}
  \caption{\textbf{Component-wise decoding analysis.} We mask either the conditioning field or the INN latent at decoding time and visualize the resulting reconstructions, revealing the distinct roles of the two paths. The masked outputs are shown only for analysis and should not be interpreted as an exact additive decomposition of the coded signal.}
  \label{Fig:component}
\end{figure}

\noindent\textbf{\textbf{Component-wise analysis.}}
To better understand the roles of the two paths, we perform a component-wise decoding analysis in \autoref{Fig:component}. Specifically, after training the full model, we mask either the conditioning field or the INN latent at decoding time and examine the resulting reconstructions. Removing the conditioning field preserves some complex local textures but severely disrupts the global content organization, consistent with the view that the conditioning path mainly provides structured spatio-temporal priors. In contrast, removing the INN latent yields outputs that show degradation patterns similar to those at low bitrates, with coarse content preserved but many fine details missing. This supports our design hypothesis that the two paths play complementary roles.

\noindent\textbf{\textbf{Complexity analysis.}}
\autoref{tab:complexity} summarizes the computational complexity of \method and the INR-based baselines. Enc/Dec FPS denotes encoding-side optimization steps and decoding-side evaluation steps per second, respectively. MACs are measured for the compression pipeline, including quantization and entropy coding. Compared with NVRC, \method uses a fixed architecture scale across different $\lambda$ values, avoiding the need to deploy separate model-size configurations for different operating points. This reduces configuration and model-management complexity, although \method is not the lightest model in terms of MACs or parameter count.

\textbf{Ablation study.}
We evaluate three degraded variants of \method: \emph{w/o masking}, which removes the proposed scheduled masking strategy; \emph{w/o modulation}, which removes the implicit conditioning field; and \emph{w/ residual}, which replaces the modulation-based interaction with direct residual coding from the conditioning-path reconstruction. As shown in \autoref{tab:ablate}, all three variants degrade rate-distortion performance. Removing masking leads to the largest BD-rate increase, suggesting that the scheduled masking strategy helps organize the latent representation for entropy coding rather than only acting as a training heuristic. Without modulation, the model loses the structured conditioning signal from the implicit path and requires noticeably more bits to reach the same reconstruction quality. The residual variant also gives higher BD-rates, indicating that directly coding the residual of the conditioning-path reconstruction is less effective than using it to modulate the invertible path. All values are BD-rate increases relative to the full model.

\begin{table}[t]
\centering

\begin{minipage}[t]{0.38\textwidth}
\centering
\small
\setlength{\tabcolsep}{4pt}
\caption{Component ablation on 720p UVG dataset.} 
\label{tab:ablate}
\resizebox{0.95\linewidth}{!}{
\begin{tabular}{rcc}
\toprule
Method & PSNR & MS-SSIM \\
\midrule
Ours w/o masking     & 85.93\% & 44.72\% \\
Ours w/o modulation  & 32.44\% & 37.17\% \\
Residual variant     & 25.81\% & 28.49\% \\
Ours                 & 0.0\% & 0.0\% \\
\bottomrule
\end{tabular}
}
\end{minipage}
\hfill
\begin{minipage}[t]{0.58\textwidth}
\centering
\small
\setlength{\tabcolsep}{4pt}
\caption{Complexity comparison on 1080p UVG dataset using one RTX 4090.}
\label{tab:complexity}
\resizebox{\linewidth}{!}{
\begin{tabular}{r|cccc}
\toprule
Method & MACs$\downarrow$ & Enc FPS$\uparrow$ & Dec FPS$\uparrow$ & Params$\downarrow$ \\
\midrule
HiNeRV & 23.0--96.0G & 20.0-37.6 & 76.7-132.1 & 0.77-3.25M \\
NVRC   & 359.6--1929.0G & 2.2-6.4 & 9.7-21.0 & 2.14-31.41M \\
Ours   & 868.82G & 3.3 & 1.8 & 15.38M \\
\bottomrule
\end{tabular}
}
\end{minipage}

\end{table}

\section{Conclusion}

We presented \method, a neural video codec designed for high-fidelity compression, where transform-induced distortion becomes increasingly important as quantization becomes mild. The codec builds on an instance-adaptive design that combines INR-style conditioning, an invertible transform backbone, and scheduled latent masking. Experiments on UVG and MCL-JCV show competitive rate-distortion performance across a wide quality range, with the strongest gains in the high-bitrate regime. On UVG, \method reduces BD-rate by 21.66\% in PSNR and 46.06\% in MS-SSIM relative to x265 in this regime, while covering more than \textbf{20 dB} in PSNR without changing the architecture scale.

Despite these encouraging results, \method remains a lossy, instance-adaptive codec. The exact invertibility of the main transform removes reconstruction error from the transform itself, but the final reconstruction is still affected by quantization and finite-precision implementation. In addition, the instance-adaptive training paradigm improves representation flexibility but incurs higher optimization cost than fully pretrained codecs. Reducing this optimization cost and extending the framework toward lossless compression are important directions for future work.


{\small
\bibliographystyle{abbrv}
\bibliography{references}

@String{Computer = "{IEEE} Computer" }

@String{Academic = "Academic Press" }

@String{Springer = "Springer-Verlag" }

@techreport{cisco2022vni,
  author       = {Cisco Systems},
  title        = {{VNI} complete forecast highlights},
  year         = {2022},
  institution  = {Cisco},
  note         = {Accessed: 2025-03-14}
}

@inproceedings{xie2021enhanced,
  title={{Enhanced Invertible Encoding for Learned Image Compression}},
  author={Xie, Yueqi and Cheng, Ka Leong and Chen, Qifeng},
  booktitle={Proceedings of the 29th ACM international conference on multimedia},
  pages={162--170},
  year={2021}
}

@article{tu2025multi,
  title={{Multi-Scale Invertible Neural Network for Wide-Range Variable-Rate Learned Image Compression}},
  author={Tu, Hanyue and Wu, Siqi and Li, Li and Zhou, Wengang and Li, Houqiang},
  journal={IEEE Transactions on Multimedia},
  year={2025},
  publisher={IEEE}
}

@article{cai2024i2c,
  title={{I2C: Invertible Continuous Codec for High-Fidelity Variable-Rate Image Compression}},
  author={Cai, Shilv and Chen, Liqun and Zhang, Zhijun and Zhao, Xiangyun and Zhou, Jiahuan and Peng, Yuxin and Yan, Luxin and Zhong, Sheng and Zou, Xu},
  journal={IEEE Transactions on Pattern Analysis and Machine Intelligence},
  volume={46},
  number={6},
  pages={4262--4279},
  year={2024},
  publisher={IEEE}
}

@article{gao2025approximately,
  title={{Approximately Invertible Neural Network for Learned Image Compression}},
  author={Gao, Yanbo and Li, Shuai and Fu, Meng and Lv, Chong and Yang, Zhiyuan and Cai, Xun and Yuan, Hui and Ye, Mao},
  journal={IEEE Transactions on Image Processing},
  year={2025},
  publisher={IEEE}
}

@inproceedings{montajabi2023invertible,
  title={{Invertible Neural Network-Based Video Compression}},
  author={Montajabi, Zahra and Ghassab, Vahid Khorasani and Bouguila, Nizar},
  booktitle={ICPRAM},
  pages={558--564},
  year={2023}
}

@article{guo2025exploring,
  title={{Exploring Invertible Encoding for Deep Video Compression}},
  author={Guo, Haifeng and Kwong, Sam and Zhou, Mingliang},
  journal={IEEE Transactions on Broadcasting},
  year={2025},
  publisher={IEEE}
}

@inproceedings{lu2019dvc,
  title={{DVC: An End-to-end Deep Video Compression Framework}},
  author={Lu, Guo and Ouyang, Wanli and Xu, Dong and Zhang, Xiaoyun and Cai, Chunlei and Gao, Zhiyong},
  booktitle={Proceedings of the IEEE/CVF conference on computer vision and pattern recognition},
  pages={11006--11015},
  year={2019}
}

@inproceedings{hu2021fvc,
  title={{FVC: A New Framework towards Deep Video Compression in Feature Space}},
  author={Hu, Zhihao and Lu, Guo and Xu, Dong},
  booktitle={Proceedings of the IEEE/CVF conference on computer vision and pattern recognition},
  pages={1502--1511},
  year={2021}
}

@article{ma2020end,
  title={{End-to-End Optimized Versatile Image Compression With Wavelet-Like Transform}},
  author={Ma, Haichuan and Liu, Dong and Yan, Ning and Li, Houqiang and Wu, Feng},
  journal={IEEE Transactions on Pattern Analysis and Machine Intelligence},
  volume={44},
  number={3},
  pages={1247--1263},
  year={2020},
  publisher={IEEE}
}

@article{ma2019iwave,
  title={{iWave}: {CNN}-based wavelet-like transform for image compression},
  author={Ma, Haichuan and Liu, Dong and Xiong, Ruiqin and Wu, Feng},
  journal={IEEE Transactions on Multimedia},
  volume={22},
  number={7},
  pages={1667--1679},
  year={2019},
  publisher={IEEE}
}

@article{kobyzev2020normalizing,
  title={{Normalizing Flows: An Introduction and Review of Current Methods}},
  author={Kobyzev, Ivan and Prince, Simon JD and Brubaker, Marcus A},
  journal={IEEE transactions on pattern analysis and machine intelligence},
  volume={43},
  number={11},
  pages={3964--3979},
  year={2020},
  publisher={IEEE}
}

@article{dinh2016density,
  title={Density estimation using real nvp},
  author={Dinh, Laurent and Sohl-Dickstein, Jascha and Bengio, Samy},
  journal={arXiv preprint arXiv:1605.08803},
  year={2016}
}

@article{helminger2020lossy,
  title={{Lossy Image Compression with Normalizing Flows}},
  author={Helminger, Leonhard and Djelouah, Abdelaziz and Gross, Markus and Schroers, Christopher},
  journal={arXiv preprint arXiv:2008.10486},
  year={2020}
}

@article{li2021deep,
  title={{Deep Contextual Video Compression}},
  author={Li, Jiahao and Li, Bin and Lu, Yan},
  journal={Advances in Neural Information Processing Systems},
  volume={34},
  pages={18114--18125},
  year={2021}
}

@inproceedings{li2023neural,
  title={{Neural Video Compression with Diverse Contexts}},
  author={Li, Jiahao and Li, Bin and Lu, Yan},
  booktitle={Proceedings of the IEEE/CVF conference on computer vision and pattern recognition},
  pages={22616--22626},
  year={2023}
}

@article{sheng2022temporal,
  title={Temporal context mining for learned video compression},
  author={Sheng, Xihua and Li, Jiahao and Li, Bin and Li, Li and Liu, Dong and Lu, Yan},
  journal={IEEE Transactions on Multimedia},
  volume={25},
  pages={7311--7322},
  year={2022},
  publisher={IEEE}
}

@inproceedings{li2024neural,
  title={{Neural Video Compression with Feature Modulation}},
  author={Li, Jiahao and Li, Bin and Lu, Yan},
  booktitle={Proceedings of the IEEE/CVF Conference on Computer Vision and Pattern Recognition},
  pages={26099--26108},
  year={2024}
}

@Conference{bjontegaard2001calculation,
  Title                    = {Calculation of average {PSNR} differences between {RD}-curves},
  Author                   = {G. Bj{\o}ntegaard},
  Booktitle                = {13th VCEG Meeting},
  Year                     = {2001},

  Address                  = {Austin, Texas, USA},
  Month                    = {April},
  Number                   = {VCEG-M33},
  Organization             = {ITU-T},

  Journal                  = {13th VCEG Meeting}
}

@inproceedings{wang2003multiscale,
  title={Multiscale structural similarity for image quality assessment},
  author={Wang, Zhou and Simoncelli, Eero P and Bovik, Alan C},
  booktitle={The thrity-seventh asilomar conference on signals, systems \& computers, 2003},
  volume={2},
  pages={1398--1402},
  year={2003},
  organization={Ieee}
}

@manual{apple2022prores,
  title        = {Apple ProRes},
  author       = {{Apple Inc.}},
  organization = {Apple Inc.},
  year         = {2022},
  month        = apr,
  note         = {White paper},
  url          = {https://www.apple.com/final-cut-pro/docs/Apple_ProRes.pdf}
}

@inproceedings{jia2025towards,
  title={{Towards Practical Real-Time Neural Video Compression}},
  author={Jia, Zhaoyang and Li, Bin and Li, Jiahao and Xie, Wenxuan and Qi, Linfeng and Li, Houqiang and Lu, Yan},
  booktitle={Proceedings of the Computer Vision and Pattern Recognition Conference},
  pages={12543--12552},
  year={2025}
}

@inproceedings{chen2023hnerv,
  title={Hnerv: A hybrid neural representation for videos},
  author={Chen, Hao and Gwilliam, Matthew and Lim, Ser-Nam and Shrivastava, Abhinav},
  booktitle={Proceedings of the IEEE/CVF Conference on Computer Vision and Pattern Recognition},
  pages={10270--10279},
  year={2023}
}

@inproceedings{li2022hybrid,
  title={Hybrid spatial-temporal entropy modelling for neural video compression},
  author={Li, Jiahao and Li, Bin and Lu, Yan},
  booktitle={Proceedings of the 30th ACM international conference on multimedia},
  pages={1503--1511},
  year={2022}
}

@inproceedings{tang2025neural,
  title={Neural video compression with context modulation},
  author={Tang, Chuanbo and Li, Zhuoyuan and Bian, Yifan and Li, Li and Liu, Dong},
  booktitle={Proceedings of the Computer Vision and Pattern Recognition Conference},
  pages={12553--12563},
  year={2025}
}

@article{kwan2023hinerv,
  title={Hinerv: Video compression with hierarchical encoding-based neural representation},
  author={Kwan, Ho Man and Gao, Ge and Zhang, Fan and Gower, Andrew and Bull, David},
  journal={Advances in Neural Information Processing Systems},
  volume={36},
  year={2023}
}

@inproceedings{kwan2024nvrc,
    author    = {Ho Man Kwan and Ge Gao and Fan Zhang and Andrew Gower and David Bull},
    title     = {{NVRC}: Neural video representation compression},
    booktitle = {Advances in Neural Information Processing Systems (NeurIPS)},
    volume    = {37},
    pages     = {132440--132462},
    year      = {2024},
    publisher = {Curran Associates, Inc.}
}

@inproceedings{maiya2023nirvana,
  title={{NIRVANA: Neural implicit representations of videos with adaptive networks and autoregressive patch-wise modeling}},
  author={Maiya, Shishira R and Girish, Sharath and Ehrlich, Max and Wang, Hanyu and Lee, Kwot Sin and Poirson, Patrick and Wu, Pengxiang and Wang, Chen and Shrivastava, Abhinav},
  booktitle={{Proceedings of the IEEE/CVF Conference on Computer Vision and Pattern Recognition}},
  pages={14378--14387},
  year={2023}
}

@article{kingma2018glow,
  title={{Glow: Generative Flow with Invertible 1×1 Convolutions}},
  author={Kingma, Durk P and Dhariwal, Prafulla},
  journal={Advances in neural information processing systems},
  volume={31},
  year={2018}
}

@article{dinh2014nice,
  title={Nice: Non-linear independent components estimation},
  author={Dinh, Laurent and Krueger, David and Bengio, Yoshua},
  journal={arXiv preprint arXiv:1410.8516},
  year={2014}
}

@inproceedings{liu2022convnet,
  title={{A ConvNet for the 2020s}},
  author={Liu, Zhuang and Mao, Hanzi and Wu, Chao-Yuan and Feichtenhofer, Christoph and Darrell, Trevor and Xie, Saining},
  booktitle={Proceedings of the IEEE/CVF conference on computer vision and pattern recognition},
  pages={11976--11986},
  year={2022}
}

@inproceedings{bai2023ps,
  title={{PS-NeRV}: Patch-wise stylized neural representations for videos},
  author={Bai, Yunpeng and Dong, Chao and Wang, Cairong and Yuan, Chun},
  booktitle={2023 IEEE International Conference on Image Processing (ICIP)},
  pages={41--45},
  year={2023},
  organization={IEEE}
}

@inproceedings{xiang2023mimt,
  title={{MIMT}: Masked Image Modeling Transformer for Video Compression},
  author={Jinxi Xiang and Kuan Tian and Jun Zhang},
  booktitle={{The Eleventh International Conference on Learning Representations}},
  year={2023}
}

@inproceedings{agustsson2020scale,
  title={Scale-space flow for end-to-end optimized video compression},
  author={Agustsson, Eirikur and Minnen, David and Johnston, Nick and Balle, Johannes and Hwang, Sung Jin and Toderici, George},
  booktitle={Proceedings of the IEEE/CVF Conference on Computer Vision and Pattern Recognition},
  pages={8503--8512},
  year={2020}
}

@article{shi2025quantizing,
  title={On quantizing neural representation for variable-rate video coding},
  author={Shi, Junqi and Chen, Zhujia and Li, Hanfei and Zhao, Qi and Lu, Ming and Chen, Tong and Ma, Zhan},
  journal={arXiv preprint arXiv:2502.11729},
  year={2025}
}

@inproceedings{jiang2025ecvc,
  title={Ecvc: Exploiting non-local correlations in multiple frames for contextual video compression},
  author={Jiang, Wei and Li, Junru and Zhang, Kai and Zhang, Li},
  booktitle={Proceedings of the Computer Vision and Pattern Recognition Conference},
  pages={7331--7341},
  year={2025}
}

@inproceedings{ho2022canf,
  title={{CANF-VC: Conditional Augmented Normalizing Flows for Video Compression}},
  author={Ho, Yung-Han and Chang, Chih-Peng and Chen, Peng-Yu and Gnutti, Alessandro and Peng, Wen-Hsiao},
  booktitle={European Conference on Computer Vision},
  pages={207--223},
  year={2022},
  organization={Springer}
}

@inproceedings{hu2020improving,
  title={Improving deep video compression by resolution-adaptive flow coding},
  author={Hu, Zhihao and Chen, Zhenghao and Xu, Dong and Lu, Guo and Ouyang, Wanli and Gu, Shuhang},
  booktitle={European Conference on Computer Vision},
  pages={193--209},
  year={2020},
  organization={Springer}
}

@article{chen2021nerv,
  title={Nerv: Neural representations for videos},
  author={Chen, Hao and He, Bo and Wang, Hanyu and Ren, Yixuan and Lim, Ser Nam and Shrivastava, Abhinav},
  journal={Advances in Neural Information Processing Systems},
  volume={34},
  pages={21557--21568},
  year={2021}
}

@inproceedings{lee2023ffnerv,
  title={{FFNeRV: Flow-guided frame-wise neural representations for videos}},
  author={Lee, Joo Chan and Rho, Daniel and Ko, Jong Hwan and Park, Eunbyung},
  booktitle={{Proceedings of the 31st ACM International Conference on Multimedia}},
  pages={7859--7870},
  year={2023}
}

@inproceedings{mercat2020uvg,
  author       = {Alexandre Mercat and
                  Marko Viitanen and
                  Jarno Vanne},
  title        = {{UVG Dataset: 50/120fps 4K Sequences for Video Codec Analysis and Development}},
  booktitle    = {MMSys},
  pages        = {297--302},
  publisher    = {{ACM}},
  year         = {2020}
}

@inproceedings{wang2016mcl,
  author       = {Haiqiang Wang and
                  Weihao Gan and
                  Sudeng Hu and
                  Joe Yuchieh Lin and
                  Lina Jin and
                  Longguang Song and
                  Ping Wang and
                  Ioannis Katsavounidis and
                  Anne Aaron and
                  C.{-}C. Jay Kuo},
  title        = {{MCL-JCV: A JND-based H.264/AVC video quality assessment dataset}},
  booktitle    = {{ICIP}},
  pages        = {1509--1513},
  publisher    = {{IEEE}},
  year         = {2016}
}

@book{bull2021intelligent,
  title={Intelligent image and video compression: communicating pictures},
  author={Bull, David and Zhang, Fan},
  year={2021},
  publisher={Academic Press}
}

@misc{x265,
    title       = {x265},
    howpublished= "\url{https://www.videolan.org/developers/x265.html}",
}

@misc{hm18,
  author= {{Joint Video Experts Team (JVET)}},
  title = {HM-18.0: HEVC Reference Software},
  year  = {2023},
  url   = {https://vcgit.hhi.fraunhofer.de/jvet/HM},
  note  = {Accessed: March 28, 2026}
}

@article{teng2024benchmarking,
  title={Benchmarking conventional and learned video codecs with a low-delay configuration},
  author={Teng, Siyue and Jiang, Yuxuan and Gao, Ge and Zhang, Fan and Davis, Thomas and Liu, Zoe and Bull, David},
  journal={arXiv preprint arXiv:2408.05042},
  year={2024}
}

@article{wiegand2003overview,
  title={{O}verview of the {H.264/AVC} video coding standard},
  author={Wiegand, Thomas and Sullivan, Gary J and Bjontegaard, Gisle and Luthra, Ajay},
  journal={IEEE Transactions on circuits and systems for video technology},
  volume={13},
  number={7},
  pages={560--576},
  year={2003},
  publisher={IEEE}
}

@article{sullivan2012overview,
  author       = {Gary J. Sullivan and
                  Jens{-}Rainer Ohm and
                  Woojin Han and
                  Thomas Wiegand},
  title        = {Overview of the High Efficiency Video Coding ({HEVC}) Standard},
  journal      = {IEEE Transactions on Circuits and Systems for Video Technology},
  volume       = {22},
  number       = {12},
  pages        = {1649--1668},
  year         = {2012}
}

@article{bross2021overview,
  title={Overview of the {Versatile Video Coding} ({VVC}) Standard and its Applications},
  author={Bross, Benjamin and Wang, Ye-Kui and Ye, Yan and Liu, Shan and Chen, Jianle and Sullivan, Gary J and Ohm, Jens-Rainer},
  journal={IEEE Transactions on Circuits and Systems for Video Technology},
  volume={31},
  number={10},
  pages={3736--3764},
  year={2021},
  publisher={IEEE}
}

@article{han2020technical,
  title={A technical overview of av1},
  author={Han, Jingning and Li, Bohan and Mukherjee, Debargha and Chiang, Ching-Han and Grange, Adrian and Chen, Cheng and Su, Hui and Parker, Sarah and Deng, Sai and Joshi, Urvang and others},
  journal={arXiv preprint arXiv:2008.06091},
  year={2020}
}

@article{chen2020overview,
  title={An overview of coding tools in {AV1}: the first video codec from the alliance for open media},
  author={Chen, Yue and Mukherjee, Debargha and Han, Jingning and Grange, Adrian and Xu, Yaowu and Parker, Sarah and Chen, Cheng and Su, Hui and Joshi, Urvang and Chiang, Ching-Han and others},
  journal={APSIPA Transactions on Signal and Information Processing},
  volume={9},
  pages={e6},
  year={2020},
  publisher={Cambridge University Press}
}

@article{hamidouche2022versatile,
  title={Versatile video coding standard: A review from coding tools to consumers deployment},
  author={Hamidouche, Wassim and Biatek, Thibaud and Abdoli, Mohsen and Fran{\c{c}}ois, Edouard and Pescador, Fernando and Radosavljevi{\'c}, Milo{\v{s}} and Menard, Daniel and Raulet, Mickael},
  journal={IEEE Consumer Electronics Magazine},
  volume={11},
  number={5},
  pages={10--24},
  year={2022},
  publisher={IEEE}
}

@article{zhao2022aom,
  title={{AOM Common Test Conditions v3. 0}},
  author={Zhao, Xin and Lei, Z and Norkin, Andrey and Daede, Thomas and Tourapis, Alexis},
  journal={Document, CWG-C038i},
  volume={5},
  year={2022}
}

@article{abdoli2024video,
  title={{Video compression beyond VVC: Quantitative analysis of intra coding tools in enhanced compression model (ECM)}},
  author={Abdoli, Mohsen and Youvalari, Ramin G and Naser, Karam and Reuz{\'e}, Kevin and L{\'e}annec, Fabrice Le},
  journal={arXiv preprint arXiv:2404.07872},
  year={2024}
}

@article{gao2025givic,
  title={{GIViC}: {G}enerative Implicit Video Compression},
  author={Gao, Ge and Teng, Siyue and Peng, Tianhao and Zhang, Fan and Bull, David},
  journal={arXiv preprint arXiv:2503.19604},
  year={2025}
}

@article{ho2021anfic,
  title={{ANFIC}: {I}mage compression using augmented normalizing flows},
  author={Ho, Yung-Han and Chan, Chih-Chun and Peng, Wen-Hsiao and Hang, Hsueh-Ming and Doma{\'n}ski, Marek},
  journal={IEEE Open Journal of Circuits and Systems},
  volume={2},
  pages={613--626},
  year={2021},
  publisher={IEEE}
}

@inproceedings{balle2017end,
  author       = {Johannes Ball{\'{e}} and
                  Valero Laparra and
                  Eero P. Simoncelli},
  title        = {End-to-end Optimized Image Compression},
  booktitle    = {{ICLR}},
  publisher    = {OpenReview.net},
  year         = {2017}
}

@inproceedings{agustsson2020universally,
  author       = {Eirikur Agustsson and
                  Lucas Theis},
  title        = {Universally Quantized Neural Compression},
  booktitle    = {NeurIPS},
  year         = {2020}
}

@inproceedings{kim2024c3,
  author       = {Hyunjik Kim and
                  Matthias Bauer and
                  Lucas Theis and
                  Jonathan Richard Schwarz and
                  Emilien Dupont},
  title        = {{C3:} High-Performance and Low-Complexity Neural Compression from
                  a Single Image or Video},
  booktitle    = {{CVPR}},
  pages        = {9347--9358},
  publisher    = {{IEEE}},
  year         = {2024}
}

@inproceedings{ladune2023cool,
  title={{COOL-CHIC}: {C}oordinate-based low complexity hierarchical image codec},
  author={Ladune, Th{\'e}o and Philippe, Pierrick and Henry, F{\'e}lix and Clare, Gordon and Leguay, Thomas},
  booktitle={Proceedings of the IEEE/CVF International Conference on Computer Vision},
  pages={13515--13522},
  year={2023}
}

@inproceedings{rippel2014learning,
  title={Learning ordered representations with nested dropout},
  author={Rippel, Oren and Gelbart, Michael and Adams, Ryan},
  booktitle={International Conference on Machine Learning},
  pages={1746--1754},
  year={2014},
  organization={PMLR}
}

@article{srivastava2014dropout,
  title={Dropout: a simple way to prevent neural networks from overfitting},
  author={Srivastava, Nitish and Hinton, Geoffrey and Krizhevsky, Alex and Sutskever, Ilya and Salakhutdinov, Ruslan},
  journal={The journal of machine learning research},
  volume={15},
  number={1},
  pages={1929--1958},
  year={2014},
  publisher={JMLR. org}
}

@inproceedings{minnen2020channel,
  title={Channel-wise autoregressive entropy models for learned image compression},
  author={Minnen, David and Singh, Saurabh},
  booktitle={2020 IEEE International Conference on Image Processing (ICIP)},
  pages={3339--3343},
  year={2020},
  organization={IEEE}
}

@article{Gao2026advances,
  title = {Advances in Neural Video Compression: A Review and Benchmarking},
  author = {Gao, Ge and Feng, Chen and Jiang, Yuxuan and Peng, Tianhao and Kwan, Ho Man and Teng, Siyue and Zeng, Chengxi and Li, Yixuan and Wang, Changqi and Hamilton, Robbie and Qi, Zihao and Zhang, Fan and Bull, David},
  journal = {Preprints 26040035},
  doi = {10.20944/preprints202604.0035.v1},
  year = {2026}
}

@inproceedings{kwan2024immersive,
  author       = {Ho Man Kwan and
                  Fan Zhang and
                  Andrew Gower and
                  David Bull},
  title        = {Immersive Video Compression Using Implicit Neural Representations},
  booktitle    = {{PCS}},
  pages        = {1--5},
  publisher    = {{IEEE}},
  year         = {2024}
}

@article{bengio2013estimating,
  title={Estimating or propagating gradients through stochastic neurons for conditional computation},
  author={Bengio, Yoshua and L{\'e}onard, Nicholas and Courville, Aaron},
  journal={arXiv preprint arXiv:1308.3432},
  year={2013}
}

@inproceedings{aaron2016pixel,
  author       = {A{\"{a}}ron van den Oord and
                  Nal Kalchbrenner and
                  Koray Kavukcuoglu},
  title        = {Pixel Recurrent Neural Networks},
  booktitle    = {{ICML}},
  series       = {{JMLR} Workshop and Conference Proceedings},
  pages        = {1747--1756},
  publisher    = {JMLR.org},
  year         = {2016}
}

@article{cayley1846quelques,
  title={Sur quelques propri{\'e}t{\'e}s des d{\'e}terminants gauches.},
  author={Cayley, Arthur},
  year={1846},
  publisher={Walter de Gruyter, Berlin/New York Berlin, New York}
}

@inproceedings{zhang2024learned,
  title={Learned rate control for frame-level adaptive neural video compression via dynamic neural network},
  author={Zhang, Chenhao and Gao, Wei},
  booktitle={European Conference on Computer Vision},
  pages={239--255},
  year={2024},
  organization={Springer}
}
}

\clearpage

\end{document}